\documentclass{article}
\usepackage[utf8x]{inputenc}
\usepackage{ICIP2021,amsmath,graphicx}
\usepackage{cite}
\usepackage{amssymb,amsfonts}
\usepackage{algorithmic}
\usepackage{textcomp}
\usepackage{xcolor}
\usepackage{caption,subcaption}
\usepackage{tikz}
\usepackage{multirow}
\usepackage{diagbox}
\usepackage{hyperref}
\usepackage{pgfplots}
\pgfplotsset{compat=1.17}
\usepackage{mathrsfs}
\usetikzlibrary{arrows}
\usetikzlibrary{positioning}
\usetikzlibrary{3d}
\hypersetup{
	colorlinks=true,
	linkcolor=blue,
	filecolor=magenta,      
	urlcolor=cyan,
	pdftitle={Self-Organized Residual Blocks for Image Super-Resolution},
	pdfauthor={O. Keleş, M. Tekalp, J. Malik, S. Kıranyaz},
	pdfstartview=FitH,
	pdffitwindow=True,
}
\graphicspath{{figures/}}

\title{SELF-ORGANIZED RESIDUAL BLOCKS FOR IMAGE SUPER-RESOLUTION }
\name{Onur Keleş$^{\star}$ \qquad A. Murat Tekalp$^{\star}$\thanks{This work was supported by TUBITAK projects 217E033 and 120C156, and a grant from Turkish Is Bank to KUIS AI Center. A. M. Tekalp also acknowledges support from Turkish Academy of Sciences (TUBA).} \qquad Junaid Malik$^{\dag}$ \qquad Serkan Kıranyaz$^{\ddag}$}

\address{$^{\star}$Dept. of Electrical \& Electronics Eng., Koç University, 34450 İstanbul, Turkey \\
	$^{\dag}$Tampere University, Tampere, Finland\\
	$^{\ddag}$Qatar University, Doha, Qatar}
\begin{document}
	%
	\maketitle
	\definecolor{col_up}{rgb}{0.8784313725490196,0.6745098039215687,0.12549019607843137}
	\definecolor{col_res}{rgb}{1,0,0}
	\definecolor{col_conv}{rgb}{0,1,1}
	\definecolor{col_img}{rgb}{0.8156862745098039,0.9411764705882353,0.7529411764705882}
	\definecolor{col_w}{rgb}{0.870588,0.796078,0.776470}
	\definecolor{col_wn}{rgb}{0.996078,0.847059,0.364706}
	\definecolor{col_out}{rgb}{1,0.5,0}
	\definecolor{col_bc}{rgb}{0,0.5,1}
	\begin{abstract}
		It has become a standard practice to use the convolutional networks (ConvNet) with RELU non-linearity in image restoration and super-resolution (SR). Although the universal approximation theorem states that a multi-layer neural network can approximate any non-linear function with the desired precision, it does not reveal the best network architecture to do so. Recently, operational neural networks (ONNs) that choose the best non-linearity from a set of alternatives, and their \textquotedblleft self-organized\textquotedblright\, variants (Self-ONN) that approximate any non-linearity via Taylor series have been proposed to address the well-known limitations and drawbacks of conventional ConvNets such as network homogeneity using only the McCulloch-Pitts neuron model. In this paper, we propose the concept of self-organized operational residual (SOR) blocks, and present hybrid network architectures combining regular residual and SOR blocks to strike a balance between the benefits of stronger non-linearity and the overall number of parameters. The experimental results demonstrate that the~proposed architectures yield performance improvements in both PSNR and perceptual metrics.
	\end{abstract}
	
	\section{Introduction}
	\label{sec:intro}
	The widely popular convolutional networks (ConvNet) \cite{lecun1998gradient} are built using the McCulloch-Pitts neuron model with the~commonly used rectified linear unit (RELU) activation. The~universal approximation theorem (UAT) \cite{funahashi1989approximate, hornik1989multilayer} states that any non-linear function can be approximated by a multi-layer network with the desired accuracy. However, it is only an existence theorem and does not provide network architectures with performance guarantees. Since the common neuron model performs only linear transformations with a simple non-linear activation, many such neurons may be needed for a sufficiently good approximation, and an architecture with a given number of layers and channels may not satisfy the performance requirements of a particular problem at hand. 
	
	To boost the learning capability with more powerful network models, operational neural networks (ONN) \cite{kiranyaz2020operational}, based on generalized operational perceptron (GOP) \cite{kiranyaz2017progressive,kiranyaz2017generalized,thanh2018progressive,tran2019heterogeneous,tran2019knowledge}, and self-organized ONNs (Self-ONNs) \cite{kiranyaz2020self,malik2020} based on generative neuron model have recently been proposed. The main idea in both models is to make the network learn the best non-linear kernel transformation. While conventional ONNs have to search for the best transformation from a pre-determined operator set library, Self-ONNs can approximate the best kernel transformation by a Taylor series expansion of the desired order. It has been noted that exhaustive search makes the use of ONNs expensive \cite{kiranyaz2020operational,kiranyaz2020self}; hence, Self-ONNs have become a more promising and computationally efficient network architecture.
	
	In this paper, we explore the use of self-organized layers for image super-resolution (SR) and compare its performance with the popular EDSR residual ConvNet architecture \cite{lim2017enhanced}. The rest of the paper is organized as follows: We review EDSR and Self-ONN architectures in Section~\ref{sec:related}. We introduce self-organized residual (SOR) blocks and network architectures with SOR blocks in Section~\ref{sec:method}. In Section~\ref{sec:eval}, we present the experimental results over the benchmark dataset. Section~\ref{sec:conclusion} concludes the paper.
	
	\section{Related Work}
	\label{sec:related}
	Self-ONN is a recent network architecture \cite{kiranyaz2020self,malik2020}. In the following, we briefly review the popular convolutional residual network architectures for image SR and the foundations of Self-ONN, which constitute the background for paper. 
	
	\subsection{Convolutional Super-Resolution Networks}
	\label{sec:convnet}
	The most popular convolutional single image SR architectures are the enhanced deep super-resolution (EDSR) \cite{lim2017enhanced} and residual channel attention networks (RCAN) \cite{RCAN2018}. The~baseline EDSR model uses 16 residual blocks \cite{he2016deep} with 64 channels each, and a pixelshuffler layer \cite{shi2016real} for upsampling. RCAN uses a residual in residual (RIR) structure, which consists of several residual groups with long skip connections. RIR allows low-frequency information to be bypassed through skip connections, making the network focus on learning high-frequency information. Although RCAN yields slightly better results compared to EDSR, in our hybrid network we employ EDSR residual blocks due to its simplicity. 
	
	\begin{figure}
		\centering
		\resizebox{0.45\textwidth}{!}
		{\begin{tikzpicture}[line join=round,>=triangle 45,x=0.75cm,y=0.75cm]
				\node[canvas is yz plane at x=0] (temp) at (0,0,0) {\includegraphics[width=4cm,height=4cm]{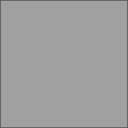}};
				\node[canvas is yz plane at x=0] (temp) at (0.25,0,0) {\includegraphics[width=4cm,height=4cm]{tensor_img.png}};
				\node[canvas is yz plane at x=0] (temp) at (0.5,0,0) {\includegraphics[width=4cm,height=4cm]{tensor_img.png}};
				
				\node[canvas is yz plane at x=0] (temp) at (2,0,0) {\includegraphics[width=4cm,height=4cm]{tensor_img.png}};
				
				\node[canvas is yz plane at x=0] (temp) at (38,0,0) {\includegraphics[width=4cm,height=4cm]{tensor_img.png}};
				
				\draw [line width=2pt] (6.5,1.5) -- (9.5,1.5) -- (9.5,-1.5) -- (6.5,-1.5) -- cycle;
				\draw [line width=2pt] (6.5,-5.5) -- (9.5,-5.5) -- (9.5,-8.5) -- (6.5,-8.5) -- cycle;
				
				
				\fill[line width=0.4pt,color=col_wn,fill=col_wn,fill opacity=0.75] (12,-4) -- (18,-4) -- (18,-10) -- (12,-10) -- cycle;
				
				\fill[line width=0.4pt,color=col_wn,fill=col_wn,fill opacity=0.75] (12,3) -- (18,3) -- (18,-3) -- (12,-3) -- cycle;
				
				\fill[line width=0.4pt,color=col_wn,fill=col_wn,fill opacity=0.75] (12,10) -- (18,10) -- (18,4) -- (12,4) -- cycle;
				
				\fill[line width=0.4pt,color=blue,fill=col_bc] (14.5,9.5) -- (17.5,9.5) -- (17.5,6.5) -- (14.5,6.5) -- cycle;
				\draw [line width=0.75pt,dash pattern=on 4pt off 4pt] (16.5,9.5)-- (16.5,6.5);
				\draw [line width=0.75pt,dash pattern=on 4pt off 4pt] (15.5,9.5)-- (15.5,6.5);
				\draw [line width=0.75pt,dash pattern=on 4pt off 4pt] (14.5,8.5)-- (17.5,8.5);
				\draw [line width=0.75pt,dash pattern=on 4pt off 4pt] (14.5,7.5)-- (17.5,7.5);
				
				\fill[line width=0.4pt,color=green,fill=green] (13,8) -- (16,8) -- (16,5) -- (13,5) -- cycle;
				\draw [line width=0.75pt,dash pattern=on 4pt off 4pt] (15,8)-- (15,5);
				\draw [line width=0.75pt,dash pattern=on 4pt off 4pt] (14,8)-- (14,5);
				\draw [line width=0.75pt,dash pattern=on 4pt off 4pt] (13,7)-- (16,7);
				\draw [line width=0.75pt,dash pattern=on 4pt off 4pt] (13,6)-- (16,6);
				
				\fill[line width=0.4pt,color=red,fill=red] (12.5,7.5) -- (15.5,7.5) -- (15.5,4.5) -- (12.5,4.5) -- cycle;
				\draw [line width=0.75pt,dash pattern=on 4pt off 4pt] (14.5,7.5)-- (14.5,4.5);
				\draw [line width=0.75pt,dash pattern=on 4pt off 4pt] (13.5,7.5)-- (13.5,4.5);
				\draw [line width=0.75pt,dash pattern=on 4pt off 4pt] (12.5,6.5)-- (15.5,6.5);
				\draw [line width=0.75pt,dash pattern=on 4pt off 4pt] (12.5,5.5)-- (15.5,5.5);
				
				\fill[line width=0.4pt,color=blue,fill=col_bc] (14.5,2.5) -- (17.5,2.5) -- (17.5,-0.5) -- (14.5,-0.5) -- cycle;
				\draw [line width=0.75pt,dash pattern=on 4pt off 4pt] (16.5,2.5)-- (16.5,-0.5);
				\draw [line width=0.75pt,dash pattern=on 4pt off 4pt] (15.5,2.5)-- (15.5,-0.5);
				\draw [line width=0.75pt,dash pattern=on 4pt off 4pt] (14.5,1.5)-- (17.5,1.5);
				\draw [line width=0.75pt,dash pattern=on 4pt off 4pt] (14.5,0.5)-- (17.5,0.5);
				
				\fill[line width=0.4pt,color=green,fill=green] (13,1) -- (16,1) -- (16,-2) -- (13,-2) -- cycle;
				\draw [line width=0.75pt,dash pattern=on 4pt off 4pt] (15,1)-- (15,-2);
				\draw [line width=0.75pt,dash pattern=on 4pt off 4pt] (14,1)-- (14,-2);
				\draw [line width=0.75pt,dash pattern=on 4pt off 4pt] (13,0)-- (16,0);
				\draw [line width=0.75pt,dash pattern=on 4pt off 4pt] (13,-1)-- (16,-1);
				
				\fill[line width=0.4pt,color=red,fill=red] (12.5,0.5) -- (15.5,0.5) -- (15.5,-2.5) -- (12.5,-2.5) -- cycle;
				\draw [line width=0.75pt,dash pattern=on 4pt off 4pt] (14.5,0.5)-- (14.5,-2.5);
				\draw [line width=0.75pt,dash pattern=on 4pt off 4pt] (13.5,0.5)-- (13.5,-2.5);
				\draw [line width=0.75pt,dash pattern=on 4pt off 4pt] (12.5,-0.5)-- (15.5,-0.5);
				\draw [line width=0.75pt,dash pattern=on 4pt off 4pt] (12.5,-1.5)-- (15.5,-1.5);
				
				\fill[line width=0.4pt,color=blue,fill=col_bc] (14.5,-4.5) -- (17.5,-4.5) -- (17.5,-7.5) -- (14.5,-7.5) -- cycle;
				\draw [line width=0.75pt,dash pattern=on 4pt off 4pt] (16.5,-4.5)-- (16.5,-7.5);
				\draw [line width=0.75pt,dash pattern=on 4pt off 4pt] (15.5,-4.5)-- (15.5,-7.5);
				\draw [line width=0.75pt,dash pattern=on 4pt off 4pt] (14.5,-5.5)-- (17.5,-5.5);
				\draw [line width=0.75pt,dash pattern=on 4pt off 4pt] (14.5,-6.5)-- (17.5,-6.5);
				
				\fill[line width=0.4pt,color=green,fill=green] (13,-6) -- (16,-6) -- (16,-9) -- (13,-9) -- cycle;
				\draw [line width=0.75pt,dash pattern=on 4pt off 4pt] (15,-6)-- (15,-9);
				\draw [line width=0.75pt,dash pattern=on 4pt off 4pt] (14,-6)-- (14,-9);
				\draw [line width=0.75pt,dash pattern=on 4pt off 4pt] (13,-7)-- (16,-7);
				\draw [line width=0.75pt,dash pattern=on 4pt off 4pt] (13,-8)-- (16,-8);
				
				\fill[line width=0.4pt,color=red,fill=red] (12.5,-6.5) -- (15.5,-6.5) -- (15.5,-9.5) -- (12.5,-9.5) -- cycle;
				\draw [line width=0.75pt,dash pattern=on 4pt off 4pt] (14.5,-6.5)-- (14.5,-9.5);
				\draw [line width=0.75pt,dash pattern=on 4pt off 4pt] (13.5,-6.5)-- (13.5,-9.5);
				\draw [line width=0.75pt,dash pattern=on 4pt off 4pt] (12.5,-7.5)-- (15.5,-7.5);
				\draw [line width=0.75pt,dash pattern=on 4pt off 4pt] (12.5,-8.5)-- (15.5,-8.5);
				
				\draw [line width=2pt] (24,0) circle (1.5cm);
				\draw [line width=2pt] (30,1.5) -- (33,1.5) -- (33,-1.5) -- (30,-1.5) -- cycle;
				
				\draw [->,line width=1pt] (2,0) -- (6.5,0);
				
				\draw [->,line width=1pt] (9.5,0) -- (12,0);
				\draw [line cap=round,line width=2pt] (5,0)-- (5,7);
				\draw [->,line width=1pt] (5,7) -- (12,7);
				\draw [line cap=round,line width=2pt] (5,0)-- (5,-7);
				\draw [->,line width=1pt] (5,-7) -- (6.5,-7);
				\draw [->,line width=1pt] (9.5,-7) -- (12,-7);
				\begin{scriptsize}
					\draw [fill=black] (5,0) circle (3pt);
				\end{scriptsize}
				
				\draw [->,line width=1pt] (18,7) -- (22.7,1.45);
				\draw [->,line width=1pt] (18,0) -- (22,0);
				\draw [->,line width=1pt] (18,-7) -- (22.7,-1.45);
				\draw [->,line width=1pt] (24,-7) -- (24,-2);
				
				\draw [->,line width=1pt] (26,0) -- (30,0);
				
				\draw [->,line width=1pt] (33,0) -- (37,0);
				
				\begin{scriptsize}
					\draw [fill=black] (0.25,0) circle (1.5pt);
					\draw [fill=black] (0.5,0) circle (1.5pt);
					\draw [fill=black] (0.75,0) circle (1.5pt);
					
					\draw [fill=black] (16.5,5.5) circle (1.5pt);
					\draw [fill=black] (16.75,5.75) circle (1.5pt);
					\draw [fill=black] (17,6) circle (1.5pt);
					
					\draw [fill=black] (16.5,-1.5) circle (1.5pt);
					\draw [fill=black] (16.75,-1.25) circle (1.5pt);
					\draw [fill=black] (17,-1) circle (1.5pt);
					
					\draw [fill=black] (16.5,-8.5) circle (1.5pt);
					\draw [fill=black] (16.75,-8.25) circle (1.5pt);
					\draw [fill=black] (17,-8) circle (1.5pt);
				\end{scriptsize}
				
				\draw [line width=4pt,dash pattern=on 8pt off 8pt] (4,11) -- (35,11) -- (35,-11) -- (4,-11) -- cycle;
				
				\draw (8,0) node[anchor=center] {\Huge $ \mathbf{(\cdot)^{2}} $};
				\draw (8,-7) node[anchor=center] {\Huge $ \mathbf{(\cdot)^{3}} $};
				
				\draw (17.25,-9.5) node[anchor=center] {\Huge $ w_{3} $};
				\draw (17.25,-2.5) node[anchor=center] {\Huge $ w_{2} $};
				\draw (17.25,4.5) node[anchor=center] {\Huge $ w_{1} $};
				\draw (24,-7.75) node[anchor=center] {\Huge $ w_{0} $};
				
				\draw (24,0) node[anchor=center] {\Huge $ \mathbf{+} $};
				
				\draw (31.5,0) node[anchor=center] {\Huge $ \sigma $};
				
				\draw (1,-4) node[anchor=center] {\Huge Input};
				\draw (1,-5) node[anchor=center] {\Huge Tensor};
				
				\draw (32,10) node[anchor=center] {\Huge Generative};
				\draw (32,9) node[anchor=center] {\Huge Neuron};
				
				\draw (38,-4) node[anchor=center] {\Huge Channel};
				
			\end{tikzpicture}
		}
		\caption{Illustration of a generative neuron for $q=3$.}
		\label{fig:gen_nrn}
	\end{figure}
	
	\subsection{Generative Neurons and Self Organized Networks}
	\label{sec:sonn}
	A Self-ONN layer is formed by generative neurons. A generative neuron approximates a non-linear function $f(x)$ by a Taylor series expansion 
	\begin{equation}
		\label{eq:TaylorSeries_f}
		f(x) = \sum_{n=0}^{\infty}\dfrac{f^{(n)}(a)}{n!}(x-a)^{n}
	\end{equation}
	around the point $ a $. If we truncate the series to $q$ terms, we~have the approximation $g(\mathbf{w},x,a)$ given by
	\begin{align}
		\label{eq:truncdTaylor}
		g(\mathbf{w},x,a) &= w_{0} + w_{1}(x-a) + \cdots + w_{q}(x-a)^{q}
	\end{align}
	where
	\begin{equation}
		\label{eq:weights}
		w_{n} = \dfrac{f^{(n)}(a)}{n!}.
	\end{equation}
	For a $c$-channel input tensor, the parameters $w_n, n=1,\dots,q$ denote $q$ banks of $c$-channel convolution kernels and $w_0$ denotes a bias. These parameters can be learned by the classical back-propagation algorithm.
	
	A generative neuron with $3\times3$ kernels, $a=0$, $q=3$, and activation function $\sigma()$ is illustrated in Figure \ref{fig:gen_nrn}. Each~neuron takes $c$-channels as input and outputs a single channel. The activation function limits outputs within a range about the~value $a$ before they are input to the~next neuron, since the~Taylor series is expanded around $a$. So, for $a=0.5$, $\sigma()$~can be taken as sigmoid bounding the output in the range~$\left[0\,\,1\right]$, or if~$a=0$, $\sigma()$ can be $\tanh(x)$ to bound the outputs in the~range~$\left[-1\,\,1\right]$. We note that if we choose $q=1$ and $a=0$, the generative neuron model reduces to the classic convolutional neuron.
	
	
	\section{Networks with Self-Organized Residual Blocks}
	\label{sec:method}
	
	We first introduce self-organized residual blocks in Section~\ref{subsec:SORB}. We then present network architectures for image SR with self-organized residual blocks in Section~\ref{subsec:architecture}. Details of training procedures are discussed in Section~\ref{subsec:traindetails}.
	\vspace{-3pt}
	
	\subsection{Self-Organized Residual Blocks}
	\label{subsec:SORB}
\begin{figure}
		\centering
		\resizebox{0.45\textwidth}{!}
		{\begin{tikzpicture}[line join=round,>=triangle 45,x=0.5cm,y=0.5cm]
				
				\fill[line width=0.4pt,color=col_conv,fill=col_conv,fill opacity=0.5] (6,6) -- (8,6) -- (8,2) -- (6,2) -- cycle;
				\fill[line width=0.4pt,color=col_out,fill=col_out,fill opacity=0.5] (9,6) -- (11,6) -- (11,2) -- (9,2) -- cycle;
				\fill[line width=0.4pt,color=col_conv,fill=col_conv,fill opacity=0.5] (12,6) -- (14,6) -- (14,2) -- (12,2) -- cycle;
				\fill[line width=0.4pt,color=col_out,fill=col_out,fill opacity=0.5] (22.5,2) -- (24.5,2) -- (24.5,-2) -- (22.5,-2) -- cycle;
				
				\draw [line width=1pt] (18,0) circle (0.75cm);
				
				\draw [->,line width=1pt] (0,0) -- (16.5,0);
				
				\begin{scriptsize}
					\draw [fill=black] (2,0) circle (1.5pt);
				\end{scriptsize}
				\draw [->,line width=1pt] (2,0) -- (6,4);
				\draw [->,line width=1pt] (8,4) -- (9,4);
				\draw [->,line width=1pt] (11,4) -- (12,4);
				\draw [->,line width=1pt] (14,4) -- (16.9383,1.0617);
				
				\draw [->,line width=1pt] (19.5,0) -- (22.5,0);
				\draw [->,line width=1pt] (24.5,0) -- (27.5,0);
				
				\draw (1,0.5) node[anchor=center] {\huge $ x $};
				
				\draw (7,4) node[anchor=center] {\rotatebox{90}{\Large  $\text{SOL} $}};
				\draw (10,4) node[anchor=center] {\rotatebox{90}{\Large  $\sigma()$}};
				\draw (13,4) node[anchor=center] {\rotatebox{90}{\Large  $\text{SOL} $}};
				
				\draw (18,0) node[anchor=center] {\Huge $ + $};
				
				\draw (23.5,0) node[anchor=center] {\rotatebox{90}{\Large  $\sigma()$}};
				
			\end{tikzpicture}
		}
		\caption{Illustration of SOR block.}
		\label{fig:SOR}
	\end{figure}
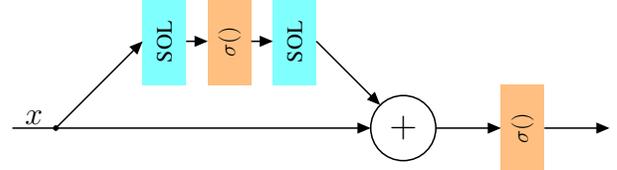

	A self-organized residual (SOR) block can be obtained by replacing all regular convolutional layers in a residual block by Self-ONN layers. In analogy with the EDSR residual blocks, we define a Self-ONN layer (SOL) without the activation function $\sigma()$ in the generative neuron to replace standard convolutional layers and use the activation function $\sigma()$ as a separate layer. Figure~\ref{fig:SOR} depicts a SOR block consisting of SOL, activation function and SOL, and another activation function after the summation. 
	
	Likewise, using SOR blocks in place of regular convolutional residual blocks, any ConvNet architecture with residual blocks, e.g., RCAN, can be transformed into a self-organized residual network architecture. The main advantage of SOR blocks over standard residual blocks is that we can obtain better performance with a fewer number of blocks, which will be demonstrated in Section~\ref{sec:eval}. \vspace{-3pt}
	
	\begin{figure*}[t]
		\centering
		\resizebox{\textwidth}{!}{
			\begin{tikzpicture}[line join=round,>=triangle 45,x=1cm,y=1cm]
				\node[canvas is xy plane at z=0] (temp) at (-14.25,0,0) {\includegraphics[width=2.4818cm,height=2cm]{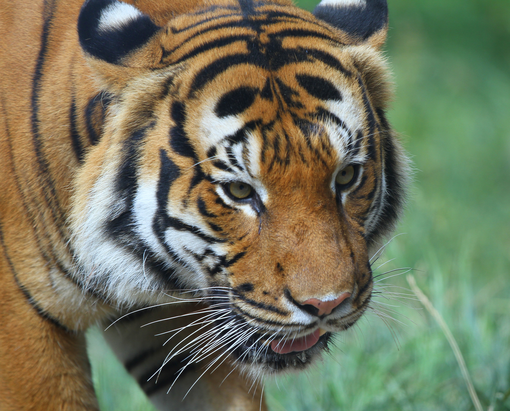}};
				\node[canvas is xy plane at z=0] (temp) at (19.5,0,0) {\includegraphics[width=4.9635cm,height=4cm]{tiger.png}};
				
				\draw [->,line width=2pt] (-13,0) -- (-11,0);
				\draw [->,line width=2pt] (-10,0) -- (-8,0);
				\draw [line cap=round,line width=2pt] (-9,0)-- (-9,3);
				\draw [line cap=round,line width=2pt] (-9,3)-- (8,3);
				\draw [->,line width=2pt] (8,3) -- (8,0.5);
				\draw [->,line width=2pt] (-7,0) -- (-6,0);
				\draw [->,line width=2pt] (-5,0) -- (-4,0);
				\draw [->,line width=2pt] (-2,0) -- (-1,0);
				\draw [->,line width=2pt] (0,0) -- (1,0);
				\draw [->,line width=2pt] (2,0) -- (3,0);
				\draw [->,line width=2pt] (4,0) -- (5,0);
				\draw [->,line width=2pt] (6,0) -- (7.5,0);
				\draw [->,line width=2pt] (8.5,0) -- (10,0);
				\draw [->,line width=2pt] (11,0) -- (12,0);
				\draw [->,line width=2pt] (13,0) -- (14,0);
				\draw [->,line width=2pt] (15,0) -- (17,0);
				
				\fill[line width=0.4pt,color=col_conv,fill=col_conv,fill opacity=0.5] (-11,2) -- (-10,2) -- (-10,-2) -- (-11,-2) -- cycle;
				\fill[line width=0.4pt,color=col_res,fill=col_res,fill opacity=0.5] (-8,2) -- (-7,2) -- (-7,-2) -- (-8,-2) -- cycle;
				\fill[line width=0.4pt,color=col_res,fill=col_res,fill opacity=0.5] (-6,2) -- (-5,2) -- (-5,-2) -- (-6,-2) -- cycle;
				\fill[line width=0.4pt,color=col_res,fill=col_res,fill opacity=0.5] (-1,2) -- (0,2) -- (0,-2) -- (-1,-2) -- cycle;
				\fill[line width=0.4pt,color=col_res,fill=col_res,fill opacity=0.5] (1,2) -- (2,2) -- (2,-2) -- (1,-2) -- cycle;
				\fill[line width=0.4pt,color=col_res,fill=col_res,fill opacity=0.5] (3,2) -- (4,2) -- (4,-2) -- (3,-2) -- cycle;
				\fill[line width=0.4pt,color=col_res,fill=col_res,fill opacity=0.5] (5,2) -- (6,2) -- (6,-2) -- (5,-2) -- cycle;
				\fill[line width=0.4pt,color=col_conv,fill=col_conv,fill opacity=0.5] (10,2) -- (11,2) -- (11,-2) -- (10,-2) -- cycle;
				\fill[line width=0.4pt,color=col_up,fill=col_up,fill opacity=0.5] (12,2) -- (13,2) -- (13,-2) -- (12,-2) -- cycle;
				\fill[line width=0.4pt,color=col_conv,fill=col_conv,fill opacity=0.5] (14,2) -- (15,2) -- (15,-2) -- (14,-2) -- cycle;
				
				\draw [line width=2pt] (8,0) circle (0.5cm);
				
				
				\draw [line width=0.2pt,color=col_conv] (-11,2)-- (-10,2);
				\draw [line width=0.2pt,color=col_conv] (-10,2)-- (-10,-2);
				\draw [line width=0.2pt,color=col_conv] (-10,-2)-- (-11,-2);
				\draw [line width=0.2pt,color=col_conv] (-11,-2)-- (-11,2);
				
				\draw [line width=0.2pt,color=col_res] (-8,2)-- (-7,2);
				\draw [line width=0.2pt,color=col_res] (-7,2)-- (-7,-2);
				\draw [line width=0.2pt,color=col_res] (-7,-2)-- (-8,-2);
				\draw [line width=0.2pt,color=col_res] (-8,-2)-- (-8,2);
				
				\draw [line width=0.2pt,color=col_res] (-6,2)-- (-5,2);
				\draw [line width=0.2pt,color=col_res] (-5,2)-- (-5,-2);
				\draw [line width=0.2pt,color=col_res] (-5,-2)-- (-6,-2);
				\draw [line width=0.2pt,color=col_res] (-6,-2)-- (-6,2);
				
				\draw [line width=0.2pt,color=col_res] (-1,2)-- (0,2);
				\draw [line width=0.2pt,color=col_res] (0,2)-- (0,-2);
				\draw [line width=0.2pt,color=col_res] (0,-2)-- (-1,-2);
				\draw [line width=0.2pt,color=col_res] (-1,-2)-- (-1,2);
				
				\draw [line width=0.2pt,color=col_res] (1,2)-- (2,2);
				\draw [line width=0.2pt,color=col_res] (2,2)-- (2,-2);
				\draw [line width=0.2pt,color=col_res] (2,-2)-- (1,-2);
				\draw [line width=0.2pt,color=col_res] (1,-2)-- (1,2);
				
				\draw [line width=0.2pt,color=col_res] (3,2)-- (4,2);
				\draw [line width=0.2pt,color=col_res] (4,2)-- (4,-2);
				\draw [line width=0.2pt,color=col_res] (4,-2)-- (3,-2);
				\draw [line width=0.2pt,color=col_res] (3,-2)-- (3,2);
				
				\draw [line width=0.2pt,color=col_res] (5,2)-- (6,2);
				\draw [line width=0.2pt,color=col_res] (6,2)-- (6,-2);
				\draw [line width=0.2pt,color=col_res] (6,-2)-- (5,-2);
				\draw [line width=0.2pt,color=col_res] (5,-2)-- (5,2);
				
				\draw [line width=0.2pt,color=col_conv] (10,2)-- (11,2);
				\draw [line width=0.2pt,color=col_conv] (11,2)-- (11,-2);
				\draw [line width=0.2pt,color=col_conv] (11,-2)-- (10,-2);
				\draw [line width=0.2pt,color=col_conv] (10,-2)-- (10,2);
				
				\draw [line width=0.2pt,color=col_up] (12,2)-- (13,2);
				\draw [line width=0.2pt,color=col_up] (13,2)-- (13,-2);
				\draw [line width=0.2pt,color=col_up] (13,-2)-- (12,-2);
				\draw [line width=0.2pt,color=col_up] (12,-2)-- (12,2);
				
				\draw [line width=0.2pt,color=col_conv] (14,2)-- (15,2);
				\draw [line width=0.2pt,color=col_conv] (15,2)-- (15,-2);
				\draw [line width=0.2pt,color=col_conv] (15,-2)-- (14,-2);
				\draw [line width=0.2pt,color=col_conv] (14,-2)-- (14,2);
				
				
				
				\draw (-10.5,0) node[anchor=center] {\rotatebox{90}{\Large  $\text{Conv,}\,3\times 3\times 64 $}};
				
				\draw (-7.5,0) node[anchor=center] {\rotatebox{90}{\Large  $\text{ResBlock-1} $}};
				\draw (-5.5,0) node[anchor=center] {\rotatebox{90}{\Large  $\text{ResBlock-2} $}};
				\draw (-0.5,0) node[anchor=center] {\rotatebox{90}{\Large  $\text{ResBlock-13} $}};
				\draw (1.5,0) node[anchor=center] {\rotatebox{90}{\Large  $\text{ResBlock-14} $}};
				\draw (3.5,0) node[anchor=center] {\rotatebox{90}{\Large  $\text{ResBlock-15} $}};
				\draw (5.5,0) node[anchor=center] {\rotatebox{90}{\Large  $\text{ResBlock-16} $}};
				
				\draw (8,0) node[anchor=center] {\Large $+$};
				
				\draw (10.5,0) node[anchor=center] {\rotatebox{90}{\Large  $\text{Conv,}\,3\times 3\times 64 $}};
				
				\draw (12.5,0) node[anchor=center] {\rotatebox{90}{\Large  $\text{Pixel-Shuffler} $}};
				
				\draw (14.5,0) node[anchor=center] {\rotatebox{90}{\Large  $\text{Conv,}\,3\times 3\times 3 $}};
				
				\draw (14,2.75) node[anchor=center] {\huge  $\text{Upsampler} $};
				
				\begin{scriptsize}
					\draw [fill=black] (-9,0) circle (2pt);
					
					\draw [fill=black] (-3.5,0) circle (1pt);
					\draw [fill=black] (-3,0) circle (1pt);
					\draw [fill=black] (-2.5,0) circle (1pt);
				\end{scriptsize}
				
				\fill[line width=2pt,dash pattern=on 1pt off 1pt,color=gray,fill=gray,fill opacity=0.2] (-1.6,2.5) -- (6.6,2.5) -- (6.6,-2.5) -- (-1.6,-2.5) -- cycle;
				
				\draw [line width=2pt,dash pattern=on 8pt off 8pt] (9,3.25) -- (16,3.25) -- (16,-2.5) -- (9,-2.5) -- cycle;
				
			\end{tikzpicture}
		}
		\caption{The EDSR network architecture. We replace only the residual blocks in the gray box with SOR blocks in the hybrid architecture. We also replace the regular convolutional layers in the Upsampler with self-organizing layers.}
		\label{fig:EDSR_arch}
	\end{figure*}
	
	\subsection{Self-Organized Residual Network Architectures}
	\label{subsec:architecture}
	We choose the EDSR baseline model, which has 16 residual blocks with 64 channels (McCulloch-Pitts neurons) each, depicted in Fig~\ref{fig:EDSR_arch}, as our starting architecture. In the following, we propose two alternative network architectures for super-resolution (SR), one using only SOR blocks and another using a combination of regular residual blocks and SOR blocks in order to evaluate their performance compared with that of the EDSR network.
	
	\subsubsection{Self-ONN with the EDSR Architecture}
	We propose to replace all EDSR residual blocks with SOR blocks, and the input and output convolution layers, including those in the upsampling layer, with self-organized layers. The main advantage of this Self-ONN is using less number of blocks than the original EDSR. The performance of the Self-ONN for different number of SOR blocks and neurons per SOL is investigated in Table~\ref{tbl:SOR_num}. 
	
	\subsubsection{Hybrid Network with EDSR and SOR Blocks}
	We next design a hybrid network with both EDSR and SOR blocks to evaluate whether combining some regular residual blocks and SOR blocks would improve performance. For this purpose, we replaced 4 of the EDSR residual blocks in Figure~\ref{fig:EDSR_arch} with SOR blocks. The evaluation of the effect of which blocks to replace is shown in Table~\ref{tbl:diff_SOR_place}.
	
	\subsection{Training Details}
	\label{subsec:traindetails}
	We use the DIV2K dataset with 800 training images as our training set \cite{Agustsson_2017_CVPR_Workshops}. We normalize input images to the~range $[0,1]$ and then subtract the mean of R, G and B channels of all images in the training set from each R, G and B image, respectively, during training and tests \cite{lim2017enhanced}.
	
	The training is iteration based, and we use the same procedure for all architectures evaluated: We take	$48\times 48$ patches from low-resolution images and their corresponding regions in high-resolution images similar to \cite{lim2017enhanced}. 
	The mini-batch size is 16. At each iteration, we select 16 images and locations of the patches within images randomly.
	In addition, each patch is subject to a random rotation of 0, 90, 180 and 270 degrees for data augmentation. 
	
	We minimize $l_{1}$ loss for 300K iterations using Adam optimizer \cite{kingma2014adam} with $\beta_{1} = 0.9$, $\beta_{2} = 0.999$ and $\epsilon=10^{-8}$. 
	The~learning rate for EDSR (baseline) is $10^{-4}$, and is halved after 200K iterations. For Self-ONNs and hybrid (EDSR + SOLs) architectures, the initial learning rate is $1.5\times10^{-4}$, and it is halved after each 100K iterations.
	
	We train $\times2$ SR networks with random initialization, while $\times4$ SR networks are trained in two alternative ways: random initialization (Random), and fine-tuning a pre-trained (Pre-T) $\times2$ SR network, except for the upsampling layer, which is randomly initialized.
	
	\section{Evaluation}
	\label{sec:eval}
	Evaluations are performed on 100 validation images from the~DIV2K dataset for both $\times2$ and $\times4$ SR with bicubic downsampling.
	
	\subsection{Evaluation of Self-ONN Hyperparameters}
	\label{subsec:exp}
	We first evaluate the effect of the number of SOR blocks and number of neurons (channels) in each SOL on the performance of Self-ONN with a base configuration, where $q=3$ for the SOR blocks and the upsampling/output layers. 

    \begin{table}[h]
        \centering
        \caption{Self-ONN performance by hyperparameters.} \vspace{-6pt}
        \begin{tabular}{|c|c|c|c|c|} 
        \hline
        \small Blocks (chan) & 4 (32) & 4 (64) & 8 (32) & 8 (64) \\\hline
        PSNR                 & 34.148 & 34.424 & 34.377 & 34.616 \\\hline
        SSIM                 & 0.9608 & 0.9624 & 0.9621 & 0.9635 \\\hline
        Parameters & \small 365,059  & \small 1,448,963 & \small 586,499 & \small 2,334,311 \\\hline
        \end{tabular}
        \label{tbl:SOR_num}
    \end{table}
	
	The quantitative performance is measured in terms of PSNR and SSIM metrics. The test results with 4 or 8 SOR blocks and 32 or 64 neurons per layer are presented in Table~\ref{tbl:SOR_num}. Clearly, the model with more SOR blocks yields better performance. The performance also increases with increasing number of neurons when the number of blocks are the same.
	
\subsection{Evaluation of Hybrid Architecture}
\label{subsec:hyb}
	In the hybrid architecture, we consider replacing some residual blocks and all upsampling/output layers in the EDSR architecture with the proposed SOR blocks and SOL, respectively. To this effect, we conduct experiments to test whether to replace the first or last 4 residual blocks with SOR blocks ($q=3$) and also whether to replace the upsampling/output layers with SOL ($q=3$) or not in the case of $\times2$ SR. 
	
	
	\begin{table}[h]
		\centering
		\caption{Searching for the best hybrid network architecture.} \vspace{-6pt}
		\begin{tabular}{|c|c|c|c|c|} 
			\hline
		SOR blocks	& \multicolumn{2}{c|}{First 4} & \multicolumn{2}{c|}{Last 4} \\\hline
		UpS/Out layers	& Conv    & SOL      & Conv   & SOL    \\\hline
		PSNR        & 34.565  &  34.612  & 34.631 & 34.658 \\\hline
		SSIM        & 0.9632  &  0.9635  & 0.9636 & 0.9638 \\\hline
		%
		\end{tabular}
		\label{tbl:diff_SOR_place}
	\end{table}
	
	The results in Table \ref{tbl:diff_SOR_place} indicate that replacing the last 4~blocks with SOR blocks and replacing the upsampling and output layers with SOL yield the best results. 
	
	\begin{table*}[t]
		\centering
		\caption{Quantitative performance evaluation of the proposed models.} \vspace{-6pt}
		\begin{tabular}{|c|c|c|c||c|c|c||c|c|c|} 
			\hline
			\multirow{3}{*}{\diagbox{Metric}{Method}} & \multicolumn{3}{c||}{EDSR (baseline)} & \multicolumn{3}{c||}{Self-ONN (8 SOR-64 + SOL UpS)} & \multicolumn{3}{c|}{EDSR-12 + 4 SOR + SOL UpS} \\\cline{2-10}
			& \multirow{2}{*}{$\times 2$} & \multicolumn{2}{c||}{$\times 4$} & \multirow{2}{*}{$\times 2$} & \multicolumn{2}{c||}{$\times 4$} & \multirow{2}{*}{$\times 2$} & \multicolumn{2}{c|}{$\times 4$}  \\\cline{3-4}\cline{6-7}\cline{9-10}
			&        		   			& Random     & Pre-T     &        & Random     & Pre-T     & 		  & Random 	   & Pre-T 	\\\hline\hline
			PSNR $\uparrow$    & 34.557 & 28.924 & 28.970 & 34.616 & 28.983 & 29.016 & 34.658 & 29.018 & 29.064 \\\hline
			SSIM $\uparrow$    & 0.9632 & 0.8857 & 0.8866 & 0.9635 & 0.8871 & 0.8875 & 0.9638 & 0.8877 & 0.8886 \\\hline
			LPIPS $\downarrow$ & 0.0896 & 0.2775 & 0.2747 & 0.0873 & 0.2742 & 0.2714 & 0.0890 & 0.2726 & 0.2694 \\\hline
		\end{tabular}  \vspace{-6pt}
		\label{tbl:results}
	\end{table*}

	\subsection{Comparative Results}
	\label{subsec:quantres}
	Based on the results in Tables~\ref{tbl:SOR_num} and~\ref{tbl:diff_SOR_place}, we compare the performances of i) Self-ONN with 8 SOR blocks and using SOL in upsampling and output layers, and ii) the hybrid architecture, where the last four blocks are SOR blocks and upsampling and output layers are SOL with that of the EDSR baseline model in Table \ref{tbl:results}. All residual/SOR blocks in all networks have 64 channels. We set $q=3$ for all SOL layers in the SOR blocks and upsampling/output layers in order to keep the number of parameters in all models approximately the same. In this study, we trained the EDSR network without using the geometric ensemble strategy adopted in \cite{lim2017enhanced}.
	
	We compare the performance of the proposed models with that of the EDSR baseline model using PSNR and SSIM measures. We also present LPIPS scores \cite{zhang2018perceptual}, which recently have been shown to correlate with human visual preferences. The arrows ($\uparrow$) and ($\downarrow$) next to the measures indicate whether a high or low value shows better performance, respectively.
	
\begin{figure}
		\centering
		\begin{tikzpicture}
			\clip(0,0) rectangle (7.175,3.5);
			\node[anchor=south west,inner sep=1] (img) at (0,0) {\includegraphics[width=0.4\textwidth]{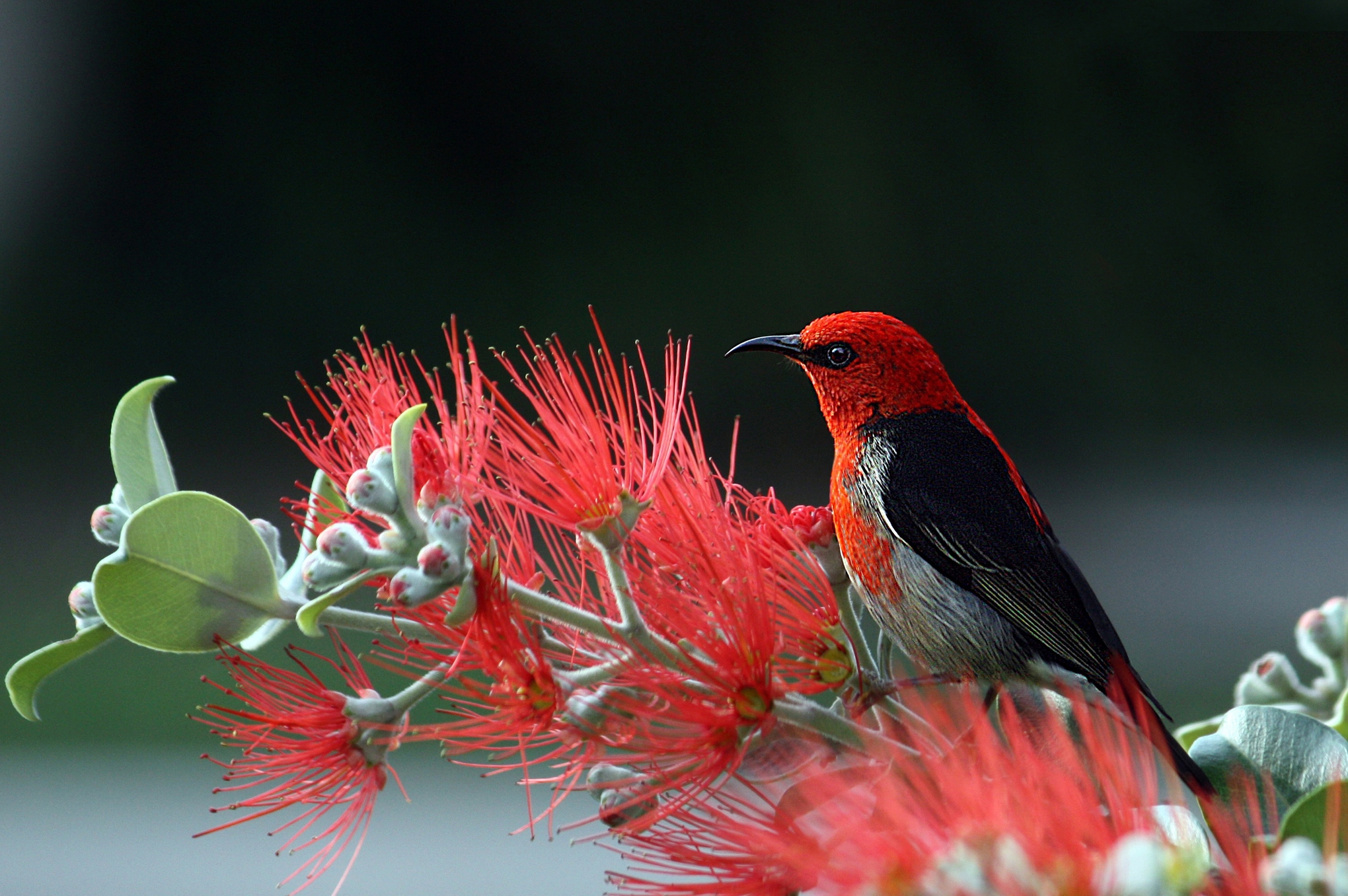}};
			\begin{scope}[x={(img.south east)},y={(img.north west)}]
				\draw[red,thick] (0.7181,0.2515) rectangle (0.8162,0.3621);
			\end{scope}
		\end{tikzpicture}
		\begin{subfigure}{0.32\columnwidth}
			\includegraphics[width=\columnwidth]{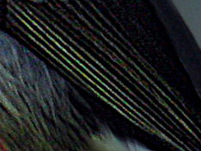}
			\caption*{Ground Truth}
			\label{fig:gt_52}
		\end{subfigure}
		\begin{subfigure}{0.32\columnwidth}
			\includegraphics[width=\columnwidth]{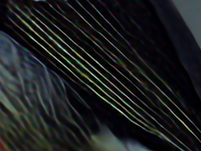}
			\caption*{Hybrid}
			\label{fig:h_52}
		\end{subfigure}
		\begin{subfigure}{0.32\columnwidth}
			\includegraphics[width=\columnwidth]{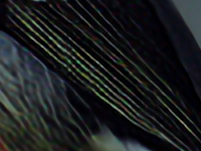}
			\caption*{EDSR}
			\label{fig:e_52}
		\end{subfigure}  \vspace{-5pt}
		\caption{Visual comparison of the EDSR baseline and hybrid network architecture with fine-tuned pre-trained model.}
		\label{fig:hybrid_vs_edsr}
	\vspace{-2pt}
	\end{figure}

\begin{figure}
		\centering
		\begin{tikzpicture}
			\clip(0,0) rectangle (7.175,3.5);
			\node[anchor=south west,inner sep=1] (img) at (0,0) {\includegraphics[width=0.4\textwidth]{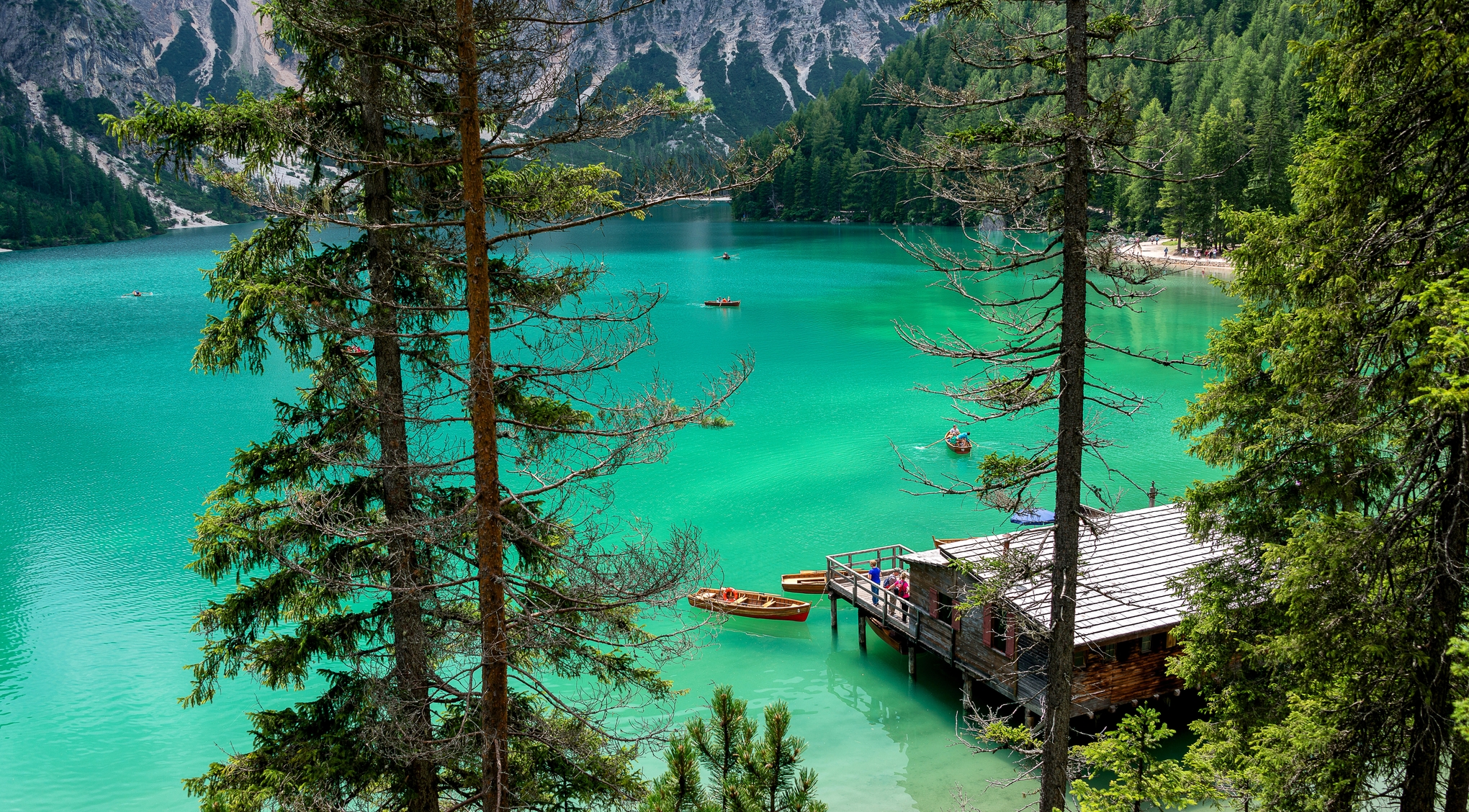}}; 
			\begin{scope}[x={(img.south east)},y={(img.north west)}]
				\draw[red,thick] (0.7181,0.3174) rectangle (0.7917,0.4060);
			\end{scope}  
		\end{tikzpicture} 
		\begin{subfigure}{0.325\columnwidth}
		\vspace*{5pt}	\includegraphics[width=\columnwidth]{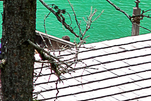}
			\caption*{Ground Truth}
			\label{fig:gt_6}
		\end{subfigure}
		\begin{subfigure}{0.325\columnwidth}
		\vspace*{5pt}	\includegraphics[width=\columnwidth]{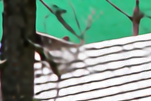}
			\caption*{Self-ONN}
			\label{fig:h_6}
		\end{subfigure}
		\begin{subfigure}{0.325\columnwidth}
		\vspace*{5pt}	\includegraphics[width=\columnwidth]{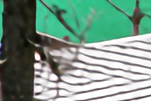}
			\caption*{EDSR}
			\label{fig:e_6}
		\end{subfigure}   \vspace{-5pt}
		\caption{Visual comparison of EDSR baseline and Self-ONN trained with random initialization for $\times4$ SR.}
		\label{fig:selfonn_vs_edsr}
	\end{figure}

	Inspection of Table \ref{tbl:results} indicates that both pure and hybrid Self-ONN models outperform the EDSR model in all quantitative metrics and in visual quality. The PSNR difference can be as much as 0.1 dB. Example visual comparison results in $\times4$ SR with randomly initialized and pre-trained models are shown in Figure~\ref{fig:selfonn_vs_edsr} and Figure~\ref{fig:hybrid_vs_edsr}, respectively. 
	We observe that the hybrid model, where we employ 12 regular residual and 4 SOR blocks and SOL for the upsampler layers, outperforms the pure Self-ONN with 8 SOR blocks although both model have approximately the same number (2.3 million for x2 SR and 2.7 million for x4 SR) of parameters.
	
	\section{Conclusion}
	\label{sec:conclusion}
	\vspace{-2pt}
	
	Self-ONNs achieve the ultimate level of network heterogeneity and expressive power whilst maximizing the network diversity along with computational efficiency, thanks to the generative neurons that have the ability to adapt the desired transformation function for each connection during training. In this paper, we propose pure and hybrid Self-ONN architectures with self-organized residual (SOR) blocks, which are composed of self-organized layers (SOL),  and show that they outperform conventional convolutional networks in the single image super-resolution task.
	
	While a SOL of order $q$ has approximately $q$ times more learnable parameters compared to a conventional convolutional block, it is important to note that a Self-ONN with 8 layers already exceeds the performance of EDSR with 16 layers. Since computations for each layer can be fully parallelized, the execution time of a network is proportional to the number of layers. Hence, Self-ONNs can achieve better performance faster than ConvNets.
	
    Our experiments demonstrate that both pure and hybrid Self-ONNs exceed the state-of-the-art SR performance of the popular EDSR network. The results also show that the hybrid model with 12 regular residual blocks and 4 SOR blocks and SOL for upsampler layers outperforms the pure Self-ONN model with 8 SOR blocks, where both models have approximately the same number of parameters.

	\clearpage
	
	\bibliographystyle{ICIP2021}
	\bibliography{ICIP2021}
	
\end{document}